\documentclass[10pt,letterpaper]{article}
\usepackage[top=0.85in,left=2.75in,footskip=0.75in]{geometry}

\usepackage{amsmath,amssymb}

\usepackage{changepage}

\usepackage[utf8x]{inputenc}

\usepackage{textcomp,marvosym}

\usepackage{cite}

\usepackage{microtype}
\DisableLigatures[f]{encoding = *, family = * }

\usepackage[table]{xcolor}

\usepackage{array}

\newcolumntype{+}{!{\vrule width 2pt}}

\newlength\savedwidth

\raggedright
\usepackage[aboveskip=1pt,labelfont=bf,labelsep=period,justification=raggedright,singlelinecheck=off]{caption}

\makeatletter
\renewcommand{\@biblabel}[1]{\quad#1.}
\makeatother

\date{}

\usepackage{lastpage,fancyhdr,graphicx}
\usepackage{epstopdf}
\fancyheadoffset[L]{2.25in}
\fancyfootoffset[L]{2.25in}

\newcommand{\added}[1]{#1}

\begin{document}
\vspace*{0.2in}

\begin{flushleft}
{\Large
\textbf\newline{Moth-inspired navigation algorithm in a turbulent odor plume from a pulsating source} 
}
\newline
\\
Alex Liberzon\textsuperscript{1*},
Kyra Harrington\textsuperscript{2},
Nimrod Daniel\textsuperscript{1},
Roi Gurka\textsuperscript{2},
Ally Harari\textsuperscript{3},
Gregory Zilman\textsuperscript{1}
\\
\bigskip
\textbf{1} School of Mechanical Engineering, Tel Aviv University, Tel-Aviv, Israel.
\\
\textbf{2} Department of Coastal and Marine Systems Science, Coastal Carolina
University, Conway, SC, USA.
\\
\textbf{3} Department of Entomology, The Volcani Center, Bet Dagan, Israel.
\\
\bigskip

%
%





* alexlib@eng.tau.ac.il

\end{flushleft}

\section*{Abstract}

Some female moths attract male moths by emitting series of pulses of pheromone filaments propagating downwind. The turbulent nature of the wind creates a complex flow environment, and causes the filaments to propagate in the form of patches with varying concentration distributions. Inspired by moth navigation capabilities, we propose a navigation strategy that enables a flier to locate a pulsating odor source in a windy environment using a single threshold-based detection sensor. The strategy is constructed based on the physical properties of the turbulent flow carrying discrete puffs of odor and does not involve learning, memory, complex decision making or statistical methods. We suggest that in turbulent plumes from a pulsating point source, an instantaneously measurable quantity referred as a ``puff crossing time'', improves the success rate as compared to the navigation strategy based on ``internal counter'' that does not use this information. Using computer simulations of fliers navigating in turbulent plumes of the pulsating point source for varying flow parameters: turbulent intensities, plume meandering and wind gusts, we obtained trajectories qualitatively resembling male moths flights towards the pheromone sources. We quantified the probability of a successful navigation as well as the flight parameters such as the time spent searching and the total flight time, with respect to different turbulent intensities, meandering or gusts. The concepts learned using this model may help to design odor-based navigation of miniature airborne autonomous vehicles.


\textit{Keywords}: \textbf{Moth navigation, chemical communication, Lagrangian model, turbulent dispersion.}



\section*{Background} \label{sec:introduction}


\added{Female arctiid moths have been shown to release pulsed pheromone signals to attract male moths~\cite{Conner1980,Schal:1987vc} whereas various species of moths have been shown to perceive pulsed pheromone signals~\cite{Kaissling:1986a,Baker1988,Grant:1997a,Rumbo1989,Bau:2005a,Bau:2002a}}. Successful mating requires the male moth to find the female within a time constraint, i.e. before the female stops emitting the pheromone and before other males find her. It is known that male moths can reach conspecific females from a long distance, overcoming obstacles such as forests or canopies \cite{Elkinton1987, Brady:1989}. Moths sense the pheromone via chemo-receptors on their antennae \cite{Schneider1964,Vickers2000,Slifer:1970uo}. If the chemical structure of the pheromone triggers the olfactory receptors in the moth antennae and is judged to be a sufficiently high quality signal~\cite{Vickers2006,Harari2011}, the male will than navigate towards the female. Along the path, the male moth has to follow the pheromone cue which is subjected to windy environment and under various turbulent conditions.


The pheromone is convected by a turbulent wind like thin, long, and convoluted filaments of varying concentration, alternated with gaps of ``clean air'' where the concentration is below the detectable limit of the male moth~\cite{Carde:1984tv,Baker1985,Mafra-Neto1994,Baker1996,Dusenbery:1989tq,Grant:1997a,Kaissling1997}. Turbulent fluctuations affect the pheromone dispersion by changing the location and increase an effective size of the concentration patch above the detectable limit~\cite{Seinfeld2006} as it would be perceived by a flying moth~\cite{Murlis1986,Murlis1992}.

It is commonly accepted that male moth strategy is based on optomotor anemotaxis. It means that moths navigate upwind (positive anemotaxis) and regulate their airspeed using visual guidance relative to its surroundings~\cite{Kennedy1974, Bau2015}. The chemotactic (using a gradient of chemical signal) navigation appears to be less likely because of the strongly intermittent and unreliable concentration signal in turbulent pheromone puffs. There is no clear gradient in the pheromone trail, in the sense that even if the mean concentration increases closer to the female, every patch that moth meets can have either increased or reduced concentration~\cite{Mafra-Neto1994}.

Another commonly observed behavior of navigating moths is the zigzagging~\cite{Kennedy1983, David1983, Mafra-Neto1994} motion. This was observed as a narrow zigzagging during the navigation to a source (a positive upwind motion) or a wide zigzagging termed ``casting'' -  a cross-wind, with a zero net upwind propagation pattern during searches of a lost signal~\cite{Kennedy1983}. The question whether this behavior is internally regulated or linked to the chemical or wind signal remains open. \added{It was suggested that the counter-turning behavior is governed by an ``internal clock''~\cite{Tobin:1981a, Baker:1997a, Carde:1997a, Vickers:1999a}, although, physical evidence for this ``clock'' was not demonstrated, to our knowledge, as yet. Other studies explained the orientation mechanism of flying moths to a pheromone sources without an internal clock  \cite{Willis1998, Preiss1986}. Internal programming was also suggested and is commonly accepted as an explanation for the moth's zigzagging angles relative to the wind direction~\cite{Kennedy1974, Kennedy1983, Carde1984, Baker1984, Willis1991}.} However, Mafra-Neto and Cardé~\cite{Mafra-Neto1996a} who studied the response of moths to single puffs of pheromone of different pulse lengths, suggested that ``single pulse reactions seem to be modified by the male's recent history of pheromone interception instead of simply being the expression of a fixed, single pulse template''.


There is a large body of literature devoted to models of moth navigation using olfactory, recently reviewed by Bau and Cardé~\cite{Bau2015}. One set of models are constructed in order to understand the behavioral strategies (for instance~\cite{Willis1991, Belanger1998, Carde2008, Grunbaum2015}), another present a general theory~\cite{Balkovsky2002, Vergassola2007}. A different type of strategies that might not be possible due to insects sensory or memory constraints, were developed for olfactory guidance of autonomous systems (for instance~\cite{Li2001,Li2010}). We review here only few works that are most relevant to the proposed navigation strategy.

We build our strategy on two common assumptions of optomotor anemotaxis: known information on local wind speed and direction and a simple chemical sensory system that provides a binary true/false input related to the detection of a pheromone signal above a certain threshold. Strategies that use similar assumptions were proposed~\cite{Willis1991, Belanger1998, Li2001, Farrell2002, Carde2008, Bau2015}. Beyond these basic assumptions the strategies are categorized according to their assumptions regarding constant/variable ground speed, internal or detection-driven anemotactic angles, plume search or acquisition and re-acquisition strategies and so forth. In this study, we suggest a navigation strategy which is similar to the so-called ``plume maintaining'' strategies~\cite{Li2001, Bau2015}.  We combined the optomotor anemotaxis with the aforementioned notion of Mafra-Neto and Cardé~\cite{Mafra-Neto1996a} regarding the ``recent history of pheromone interception'' in order to develop a robust and feasible algorithm of a flier olfactory navigation in turbulent patchy plumes. This strategy does not require memory or a complex decision making mechanism, uses a single threshold-based binary odor sensor, and it is based on an instantaneously measured quantity of a ``puff crossing time''. 

The paper is organized as follows. First we describe the components used in our simulations: a commonly used wind model~\cite{Farrell2002}, which comprises of a mean wind, turbulent fluctuations, meandering and gusts, and the Lagrangian puff dispersion model~\cite{Zannetti1981, Boeker:2011}.  Then we present the navigation algorithm, followed by the summary of the simulation results and conclusions.

\section*{Simulations}

In order to test the efficiency of the proposed strategy we implement numerical simulations based on two models: the odor dispersion in the atmosphere emitted from a pulsating point source and a algorithm used by a flier to navigate to the source. First, we simulated the odor dispersion in the Lagrangian framework. The odor patches are released from a point source at a constant rate. The patches are convected and dispersed in space and time due to the mean flow and we simulated the effects of turbulence, meandering and gusts as described below. Multiple fliers at rest are positioned at random locations downstream from the source. When one of the dispersed patches arrives at a flier location the navigation simulation starts according to the algorithm described in the following section.     

\subsection*{Lagrangian model of puff dispersion\label{sub:Discrete-puffs-model}}

\begin{figure}[!ht]
\includegraphics[width=1\textwidth]{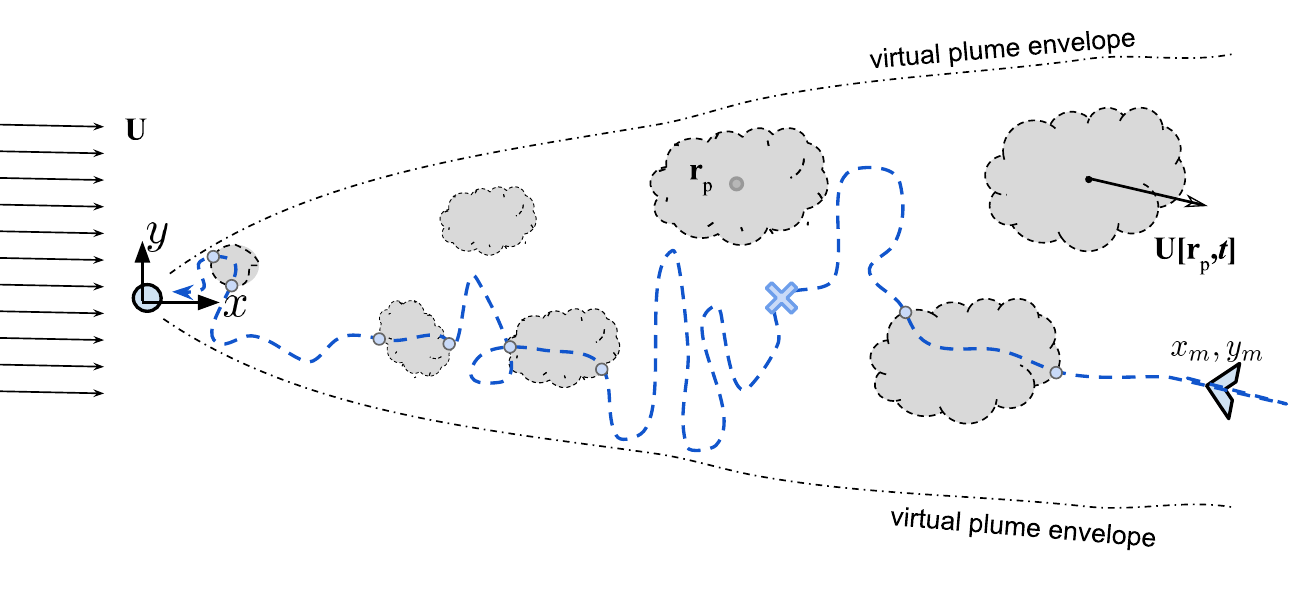}
\caption{{\bf A sketch of a series of discrete puffs and the trajectories
of a male moth (view from above)}. A female moth is denoted as a small circle
in the origin. The lowest detectable concentration of each puff is
marked by a dashed contour line. A dash-dotted envelope denotes the
limits of an average long term concentration distribution of a virtual
plume. This figure manifests the major feature of the patchy plume
- the size of patches of pheromone, the distance between them and
\added{the width of an isoline of the long time averaged concentration (i.e. ``virtual plume'')} grow proportionally in any given turbulent
flow. The cross symbol $\times$ denotes schematically the turning point where the
new casting search starts. The lateral spread of the search is equivalent
to the size of the last patch, marked by two small circles on the moth
path line. \label{fig:sketch}}
\end{figure}

Odor puff dispersion from a pulsating point source is a solution of the advection-diffusion equation~\cite{Boeker:2011}, formulated in Eq.~\ref{eq:series}. We wrote the equation in the form that reflects the pulsed type of release: each pulse releases a concentrated odor that moves downstream while dispersing, resulting in the formation of a cloud-like pattern named in the literature as ``puffs''. All the puffs together averaged for a very long time $t \to \infty$ are described through the average concentration distribution in space and time, $\overline{C}(\mathbf{r},t)$, as a function of the instantaneous wind velocity (embedded in positions of puffs $\boldsymbol{r}_p$) and turbulent fluctuations (that determine the puff sizes through the values of $\sigma_{h,z}$):
\begin{equation}
\overline{C}(\mathbf{r},t)=\frac{m}{(2\pi)^{3/2}\sigma_{h}^{2}\sigma_{z}}\sum_{i}^{\infty}H(t-t_{i})\exp\left[-\frac{| \mathbf{r}_{p}(t-t_{i})|^{2}}{2\sigma_{h}^{2}}-\frac{z_{0}}{2\sigma_{z}^{2}}\right] \label{eq:series}
\end{equation}
\noindent where $m$ is the amount of released odor, $\sigma_{h}$ and $\sigma_{z}$ are spatial variances of the concentration field distribution, $\boldsymbol{r}_{p}$ is the location of the puff center of mass which is an integral of the wind velocity along puff trajectory, $t_{i}$ is the time step in the simulation, $z_0$ is the release height above the ground, and $H(t-t_{i})$ is a Heaviside step function. The initial conditions are as follows: the odor pulsating source is located on the ground at origin $O$ on a fixed Cartesian coordinate system $Oxyz$, with the axis $Oz$ directing upward (against gravity), the axis $Ox$ pointing downwind (in the direction of the mean wind), and the axis $Oy$ corresponds to the spanwise or transverse direction (Fig.~\ref{fig:sketch}). A ``virtual plume envelope'' in Fig.~\ref{fig:sketch} is an isoline of the field at some arbitrary low threshold of concentration, $\overline{C}(\mathbf{r},t) = C^*$, for instance at the level of detectable concentration of the moth sensory system, $C^*$. Only a stationary observer that measures an intermittent concentration field of randomly passing puffs for an infinitely long time could recover the position of this ``edge of a plume''. Otherwise, this information is not available for the flier that needs to find a source at a finite time and can only sample the intermittent odor information at random positions along its path as we depict schematically by dots along a convoluted path representing a motion of the flier in Fig.~\ref{fig:sketch}. 
%

For the sake of clarity we address here only the simplest case of a constant release rate, with time intervals $T$, which is also supported by evidence of female moth pheromone release rates \cite{Bau2015,Conner1980,Schal:1987vc}. Furthermore, since the duration of each puff released was observed to be about several microseconds~\cite{Conner1980}, in our model each release represents an instantaneous puff of the equivalent amount of odor $m$. The concentration $C({\bf r},t)$ of the odor in the atmosphere (which is equivalent to mass per unit volume) depends on the coordinates ${\bf r}=(x,\,y,\,z)$ and the time; the average concentration level in the puff decreases as the puff moves downstream because puff size increases. The location of the center of the odor puff ${\bf r}_{p}=[x_{p}(t),\,y_{p}(t),\,z_{p}(t)]$, is considered to be a function of time. Following \cite{Zannetti1981}, we describe the motion of the puff center:
\begin{equation}\label{eq:r_p}
{\bf r}_{p}(t+\Delta t)={\bf r}_{p}(t)+{\bf U}[\mathbf{r}_{p},t]\Delta t
\end{equation}
\noindent where $\Delta t$ is the time increment and ${\bf U}({\bf r},t)$ is the instantaneous wind velocity. The spatial variances of the concentration distribution, $\sigma_{h,z}$, depend on the turbulent state of the flow and the environmental conditions that may influence the distribution~\cite{Zannetti1981}. The variances $\sigma_{h,z}$ presented in Eq.~\ref{eq:series} depend on the puff's age as $\sigma_{h,z}=A_{h,z}(t-t_{i})]^{B_{h,z}}$, where $A_{h,z}$ and $B_{h,z}$ are constants depending on the atmospheric conditions and Pasquill-Gifford atmospheric stability classes. It can be assumed for practical purposes that $B_{h,z} \approx 1$~\cite{Boeker:2011,Turner1971}. 
The constant coefficient in this model represents the turbulent intensity that effectively defines the Gaussian variances, $\sigma_{h,z}$, responsible for the decrease of concentration of puffs, increase of their size and rate of spread as they move with the flow. 

\added{Turbulent wind velocity ${\bf U}[\mathbf{r}_{p},t]$ in this problem is defined as a two-dimensional, two-component vector field, $u(x,y,t), v(x,y,t)$. We model it using the following components decomposed according to their relevant time scales. The word ``relevant'' means in respect to the time scales of a flier, where the shortest time scale is related to the modeled ground speed (order of magnitude of seconds, $\sim \mathcal{O}(1)$ sec) and the longest relevant time scale to the average time to find the source (order of magnitude of hundreds of seconds, $\sim \mathcal{O}(100)$ sec):
\begin{eqnarray}
u(x,y,t) & =  & U + U_{\text{g}}(t) + u'(x,y,t) \\ \nonumber
v(x,y,t) & = &  V_{\text{m}}(t) + v'(x,y,t)
\end{eqnarray}
\begin{enumerate}
\def\labelenumi{\arabic{enumi}.}
\item $U$ is the mean wind speed in the streamwise direction (upwind, downwind) and it is not changing its direction or its amplitude during the course of the simulation. This means that it is a property of the flow that can be changed only on a much slower time scale than the relevant time scale of a flier to reach the source, in other words between different runs of the simulation. Note that the mean flow is defined as uni-directional, i.e. $V = 0$. 
\item $u', v'$ - turbulent fluctuations, modeled here as two different random processes (uncorrelated, varying at every location and at every time step) which are much faster than the flier motion. It means that at every step a flier will encounter a new random wind speed at a new random direction. The time history of turbulent fluctuations defines the spread of a pheromone puff and its size at different locations, relative to the mean wind speed, $U$. Therefore the key quantity is not an amplitude of fluctuations, but a ratio of the root-mean-square of turbulent fluctuations to the mean wind, $u'_{\text{rms}}/U$. For the sake of simplicity in our model we keep $u'$ and $v'$ at the same level of their root-mean-square values, i.e. same turbulent intensity in both directions. 
\item $U_{\text{g}}(t)$ is an intermittent change of the mean wind speed (the flow changes rapidly everywhere in the flow and the amplitude can be changed up to $\pm 50\%$ of the original mean wind speed) that is modeled as a series of  steps that appear only few times during the simulation (at random), hold for an order of tens of seconds, $\sim \mathcal{O}(10)$ sec). Gusts are ``sudden increases of the mean wind speed'' and therefore $U_{\text{g}}(t)$ modeled as a function of time that modifies the streamwise velocity. Indeed this property of the flow leads to intermittent gaps in distance between the groups of puffs (but not between neighbor puffs that are driven by turbulent intensity). We show below how gusts can affect the success rate of navigation strategies.
\item $V_{\text{m}}(t)$ is a so-called meandering, which is a slow modulation (on the order of hundreds of seconds, $\sim \mathcal{O}(100)$ sec) of the direction of the mean wind speed. It is modeled as a periodic change of the flow direction everywhere in the field and it affects the long-term average path of the plume centerline, as if it is visible to a steady observer. From a perspective of a flyer, meandering is not different than the turbulent fluctuations, but the slow drift of a plume centerline from its direction at the moment of puff releases does lead to a more frequent loss of the plume.
\end{enumerate}
}

\medskip

\begin{figure}[ht]
\centering \includegraphics[width=.8\columnwidth]{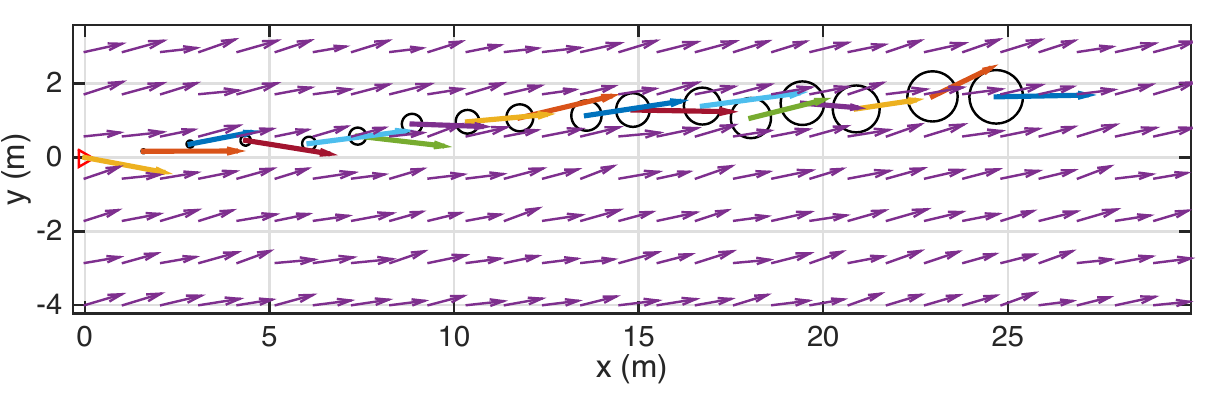} 
\caption{{\bf Typical velocity field and puff locations.} Arrows -- instantaneous two dimensional
velocity field $\mathbf{U}(x,y)$; circles -- puff locations (symbols size is arbitrary and not related to the puff size). Velocity
field is a sum of the wind velocity and random, turbulent fluctuations,
defined by turbulent intensity. \label{fig:simulation_result}}
\end{figure}

Utilizing the described models, the advection of the emitted odor is completely resolved: the average odor concentration, as shown in Eq.~\ref{eq:series}, is due to the series of puffs released at time instants $t_{i}\;(i=1,2,\ldots)$ up to the time instant, $t$, and their center positions are described by the Eq.~\ref{eq:r_p}. This solution is very similar to the commonly accepted one~\cite{Farrell2002}. A typical velocity field is shown as arrows together with the advected puffs shown schematically as circles in Fig.~\ref{fig:simulation_result}.


\subsection*{Navigation strategy algorithm \label{sub:Navigation-model}}

We model a self-propelled navigator flying at airspeed ${\bf \mathbf{V}}$  larger than the mean wind speed such that it keeps a constant ground speed. The goal of the navigator is to reach the pulsating source, in our case at the origin $O$. Arrival at a nearby location within a certain radius $R$ around the origin is defined here as a ``successful search''. To locate the pulsating source, the navigator is capable of changing the upwind angle, defined with respect to the instantaneous wind direction at the position of the navigator. To this end, the assumptions are similar to the previous works based on optomotor anemotaxis~\cite{Li2001,Bau2015}. The ratio between the number of the navigator successful searches and the total number of navigation attempts defines the probability of a successful search of a given algorithm under certain conditions of mean wind speed, meandering, gusts, turbulent intensity, and initial positions. \added{Note that we do not count the total number of fliers, as some will never meet a first puff and will not initiate a navigation attempt}. The navigation strategy is developed according to these principles:
\begin{enumerate}
\item A moth-inspired flier is insensitive to the odor if its present concentration is below a lowest detectable level.  
\item The flier estimates can steer at the same angle relative to the ground.
\item At every instant an odor patch is detected, the navigator moves upwind. Upwind means that the flier steers at 180 degrees in respect to the instantaneous wind speed vector.
\item During counter-turns, the angles $\alpha_{s}$ are selected randomly \added{within a range between 30 and 120 degrees: $\alpha_{\text{min}} = 30^\circ, \; \alpha_{\text{max}} = 120^\circ $. The values are supported by flight experiments in moths~\cite{Kennedy1974,David1983,Mafra-Neto1994}. At 90$^\circ$ the counter-turns yield the zero net upwind propagation and at angles above 90 degrees the flier drifts backwards (downwind) in respect to the ground.} The angles are kept constant during the counter-turns within the same cycle (until the next crossing event). Therefore, in some cases it also moves downwind due to the instantaneous variations of the wind at the position of the flier.
\added{It is noteworthy that these assumptions are due to the optomotor
anemotaxis~\cite{Kennedy1974}, supported by experiments studying moth navigation ~\cite{Kaissling1997,Carde2008}}.
\item The core principle which is central to this work is that the flier crosses the pheromone puffs (i.e. areas above the detectable level of odor concentration) in a finite time. The time interval that takes a flier to cross the last odor patch, defined here as a ``puff crossing time'', $t_{c}$. \added{This is the key quantity in the proposed navigation strategy and it is assumed to be measured locally at every crossing event}. 

\end{enumerate}

\begin{figure}[!ht]
\centering\includegraphics[width=.8\columnwidth]{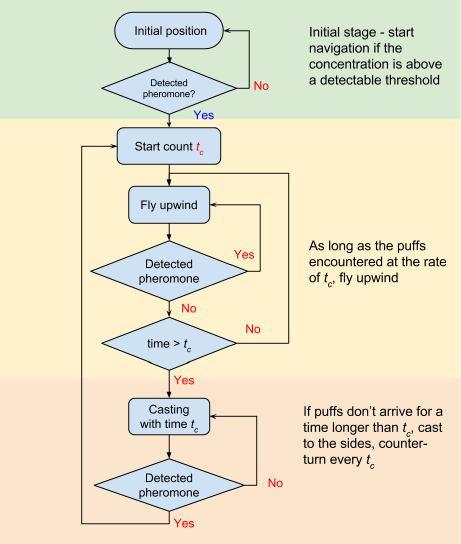}
\caption{ {\bf Flow chart of the search navigation algorithm.} \label{fig:Block-diagram}}
\end{figure}

The proposed navigation model can be described through a flow chart, as shown in Fig.~\ref{fig:Block-diagram}. 
%

A brief description of the set of actions suggested in Fig.~\ref{fig:Block-diagram} is given herein: 
\begin{enumerate}
\item A flier starts moving if an odor patch at concentration above a detectable threshold crosses its initial position at some point of time. \item Time of the patch to cross the initial position is detected as $t_c$.  
\item The flight upon every detection is upwind. 
\begin{enumerate}
\item The upwind continues for $t_c$ (at a constant ground speed it creates an upwind surge which is proportional to the size of the last puff as it was observed by Ref.~\cite{Mafra-Neto1996a})
\item \added{If no new odor is detected within $t_c$, the counter-turning starts. Each counter-turn motion lasts for $t_c$. Each turn is at some random angle in the range between  $\alpha_{\text{min}} = 30^\circ$ and $\alpha_{\text{max}} = 120^\circ $. At constant ground speed the counter-turns that last $t_c$ for every turn create a constant width proportional to the puff size. Recalling the odor dispersion model in Eq.~\ref{eq:series} one can understand that this is advantageous to search for the new puff within a limited region that is decreasing as the flier propagates towards the source.}
\item If a new odor is detected, the new $t_c$ is measured and the flowchart goes back to point 2, flying upwind (point 3). 
\end{enumerate}
\end{enumerate}

We do not set a limit on the overall time of flight of fliers within the simulation. Instead, we record and analyze the flight parameters and performance metrics of all the flights produced as a result of the simulated navigation algorithm. Flights are then marked as successful or unsuccessful depending solely on their crossing point with the source position line $(x=0,y)$ with respect to a predefined radius, $R$.

\section*{Results and discussion}

The aim of our work was to develop a bio-inspired algorithm to search for a pulsating source of an odor (i.e. chemical substance) in a turbulent environment. A patchy turbulent plume containing odor parcels is simulated as described by the Lagrangian model in Eq.~\ref{eq:series}. The trajectories (flight paths) of numerous fliers in generated turbulent fields of odor parcels are calculated using the developed navigation algorithm, presented in Fig.~\ref{fig:Block-diagram}.

The proposed model depends on a number of biotic and abiotic variables that will eventually influence the probability of success of the navigator. The \added{free} variables are: wind speed with meandering and gusts, turbulent intensity, emission rate, emitted mass, initial position of the flier, a threshold of the chemical sensor, counter-turn angles and others. \added{For the present set of simulations we show the results for a set of constant wind speed, emission rate and mass. We first describe a typical result of the simulation followed by a sensitivity test for the success probability to changes in the turbulent intensity, and then proceed to the effects of meandering, gusts and compare our strategy with the pre-defined constant ``internal counter'' model. }

\subsection*{Simulation parameters}\label{sec:simulation}

Using a set of typical parameters \added{(mean
wind speed was set at 1 m/s, the flier velocity ground speed was kept
constant and equal to 0.3 m/s)} we generate a turbulent field, initiate the release of a pulsating source and locate fliers randomly in the field. After a certain time that takes the odor patches to cross the flier position (see for instance Fig.~\ref{fig:simulation_result}), the flier navigation trajectories are generated. The odor pulsating source was located at the origin $(x,y) = (0,0)$. The initial locations of the fliers were spread randomly between 6 and 20 m away from the odor source (in the downstream direction) and between $\pm 0.4 - 1$ m transversely for different wind parameters. This range was chosen to match with the mean wind speed, meandering amplitude and the ``virtual plume'' width. Naturally, the fliers originally positioned far away downstream would encounter puffs at a much later time, and if positioned at large transverse distance might never encounter a puff. This is a practical constraint that is necessary in order to run multiple simulations to increase statistics. From a moth-inspired point of view we can see this constraint as an analogy to biological constraints: search time, competition, or energy expenditure on male moths searching for a female moth in a turbulent environment. 

We vary the turbulent intensity from 5\% to 30\% in increments of 5\%. This choice of the intensities range is aimed to simulate typical turbulent levels and conditions in the atmosphere~\cite{Boeker:2011,Turner1971}. The dispersion is set through an independent parameter, variance $\sigma_h$ which controls the rate of the patch growth and proportional to turbulent intensity. Consequently, these parameters will affect the size distribution of the puffs, and, eventually the flier chance to reach the source. At higher turbulent intensity the flier will encounter larger areas that it can navigate through, as well as the number of detections and $t_c$ values that will increase. Note that at the same distance from the source, the low level of turbulence generates a relatively smaller regions covered by odor puffs and therefore a smaller \added{width of the long time averaged concentration (``virtual plume'')} as compared to higher turbulence intensity levels.


\subsection*{Performance analysis and metrics}

\begin{figure}[!ht]
\centering\includegraphics[width=0.74\textwidth]{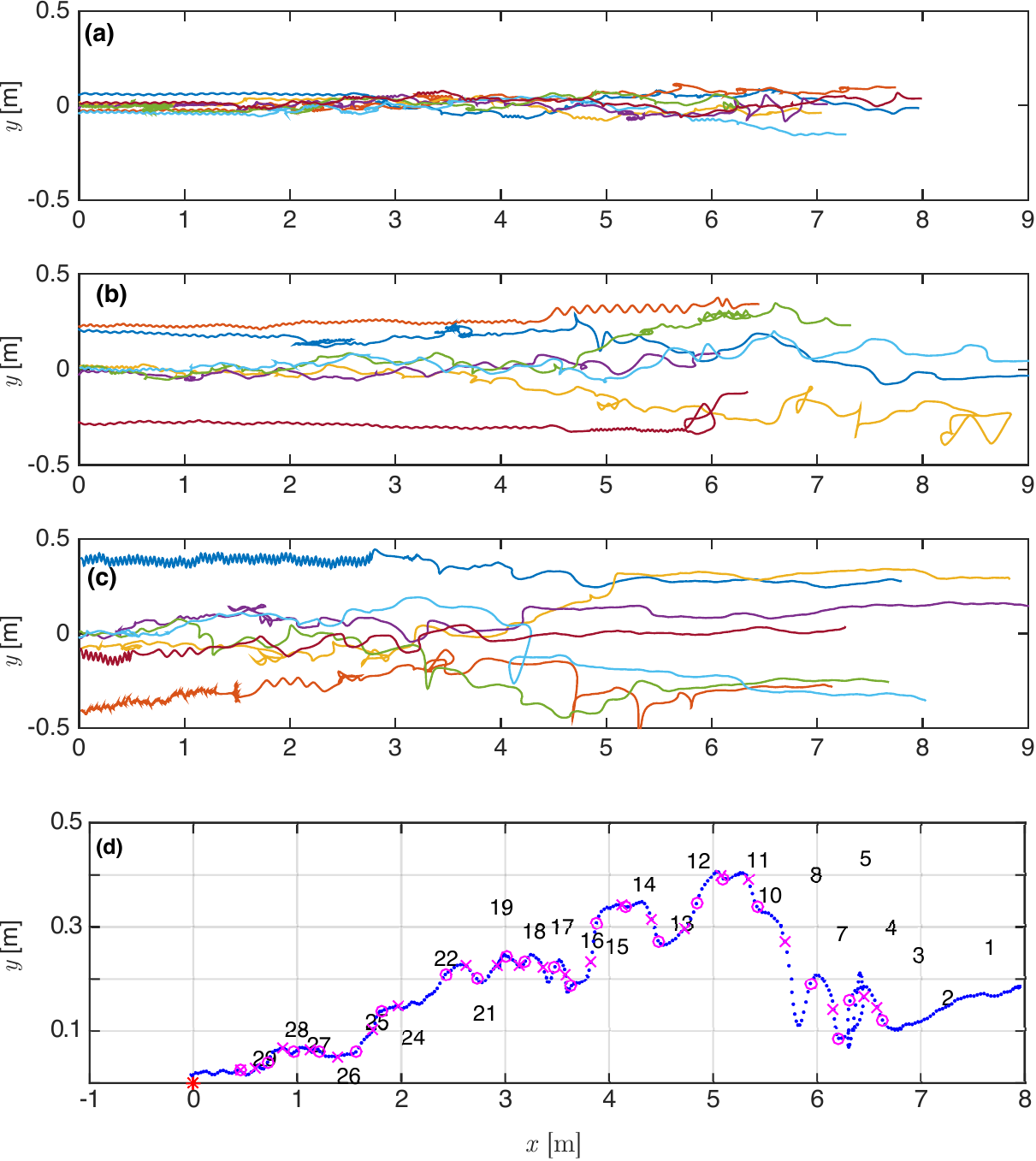} 
\caption{{\bf Typical paths of a navigator for different turbulent intensities and
representative sets of flier coordinates $(x_{m},y_{m})$, $\alpha_{s}\in$
($30^{\circ}-120^{\circ}$)}. (a) low 5\% turbulence intensity; 
(b) 15\% intermediate turbulence intensity;
(c) 30\% high turbulence intensity. \added{(d) A typical flight path with symbols showing the detection points of puffs and their identification number.} \label{fig:turbulence_intensity} }
\end{figure}

A total of 1000 simulation runs have been performed for each turbulent intensity level in the range of 5\% - 30\%. To minimize any possible bias due to a specific turbulent field simulation, we create 10 different turbulence/plume fields and ran 100 navigation simulation repetitions per field. Fig.~\ref{fig:turbulence_intensity} depicts representative examples of flier trajectories as a function of turbulent intensity. Subplots a, b and c correspond to 5, 15 and 30\%, respectively. For the sake of clarity we plot only 7 randomly selected flights for each turbulence intensity, to serve as a representative group. Each navigator flight varies for every run of the turbulence field/odor simulation, however, these results are consistent with those flights observed over the complete data set. \added{Note that we do not include in our statistics fliers that have
not started the flight during the simulation run time. The flight time
is recorded from the moment the flier start flying until it crosses the
line of $x=0$ that corresponds to the location of the odor source ($x,y =
0,0$). Once the moth passes this line, regardless of the transverse
location ($y$), the time is determined. The transverse location in respect to the radius of the detection zone defines if the flight is successful or not.} Fig.~\ref{fig:turbulence_intensity}d demonstrates in details a typical path with markers of the places at which the puffs were crossed and decisions regarding the following navigation made. 

%

Comparing the subplots we observe that the initial path of moths are closest to a straight line at the lowest turbulence intensity represented (5\%). There is an apparent increased variance with increasing turbulence intensities (15 and 30\%). The flier moves upwind after initial contact with the odor patch, and continues to fly upwind for a timed period equal to the last puff crossing time. Once the time that corresponds to upwind flight is larger than the time spent crossing the last odor patch, the flier begins counter-turning; this behavior is denoted by the stretches of zigzagging flights. It can be observed that fliers flying in low turbulence (5\%) flow are less successful reaching the source, ending at distances larger than a prescribed radius from the source. In a higher turbulent intensity the spatial area of patches are larger, increasing the likelihood of contact. Unsuccessful flights occur (at different ratio) for all turbulent intensity values tested, and are visible in Fig.~\ref{fig:turbulence_intensity} as flights that crossed at large transverse distance from the source, $y > R$.

\begin{figure}[!ht]
\centering \includegraphics[width=.7\textwidth]{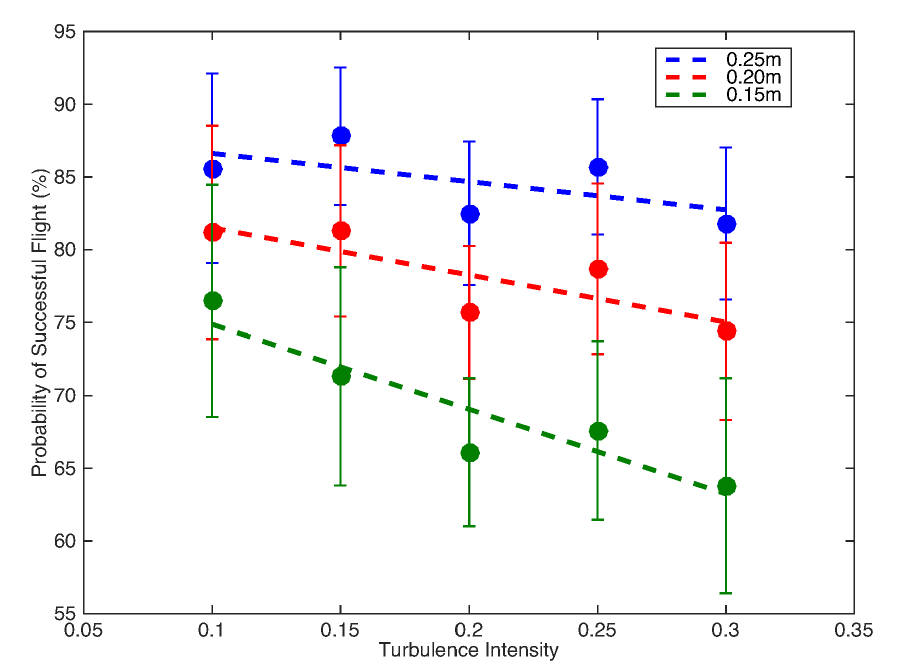} 
\caption{{\bf Probability of success as a function of increasing turbulent intensity for three different radii around the pheromone source.} Error bars demonstrate the standard deviation of 1000 simulation runs for random initial positions and simulated turbulent flow fields. \label{fig:probabilityradius} }
\end{figure}

We test the performance of the navigation strategy by changing the turbulent intensity and summarize the outcomes using the probability of success and the measure of energy expenditure, as shown in Fig.~\ref{fig:probabilityradius}. In order to remove a bias due to a wider virtual plume width at higher turbulence levels, we also vary the radius of arrival and demonstrate the results for three radii (0.15, 0.20, and 0.25 m). These values are arbitrarily chosen with some resemblance to the possible range of visual cues and to the empirical evidence that a commonly cited success rate of male moth flights in wind tunnel experiments is about 80\%~\cite{Kennedy1974,Kaissling1997,Balkovsky2002,Carde2008}.


Fig.~\ref{fig:probabilityradius} displays the probability of a successful flight for the different turbulent intensities and the three radii around the origin. The groups of simulations are used to define the average value (shown as symbols) and the standard deviation (error bars) of each group of simulations. 

The smaller is the radius chosen the success rate is also decreased. However, it is important to note that the trends are similar for all radii and therefore are not related to the distribution of initial location of the fliers. The repetitive trend that is a key result in this work is the decreased probability of success for increasing turbulent intensity. Within a given radius, the values are decreased with increasing turbulence intensity; this trend is present in all three radii used, however it is more prominent within the smallest radius. The decreased probability of success associated with increased turbulence can be explained, in part, by increased mixing within the flow, and the increased variability of signals the flier is perceiving.

\begin{figure}[!ht]
\centering\includegraphics[width=.8\textwidth]{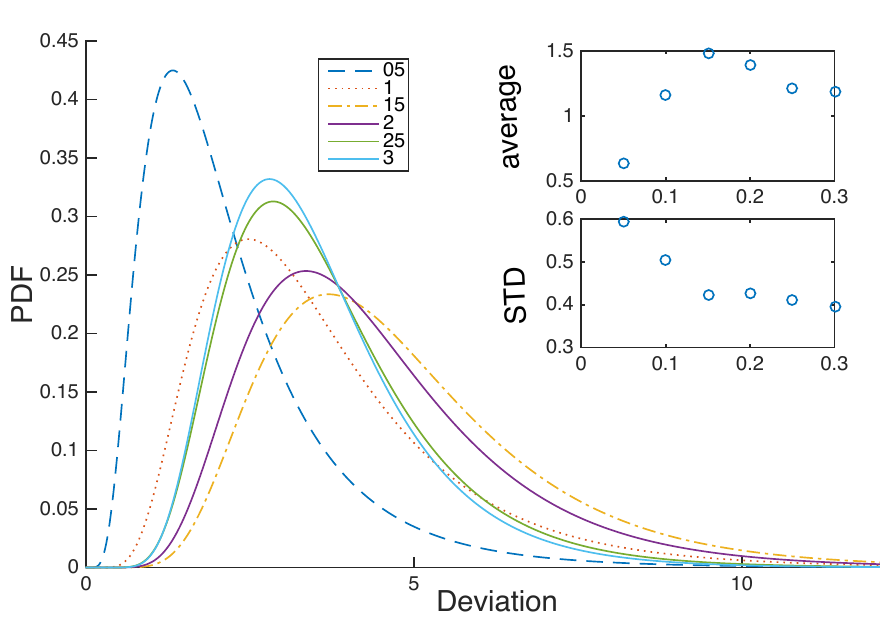} 
\caption{{\bf Probability density function of lateral deviation} (defined as a sum of all lateral deviations from a straight line between the first and last trajectory points, normalized by the length of the straight line) for different values of turbulent intensity shown in the legend. Insets shown the average values and the standard deviations of the distributions as a function of turbulent intensity.\label{fig:deviations} }
\end{figure}

In addition to the probability of success we explore other metrics of the model performance that relate to the cost of search with respect to turbulent intensity. First, we use the ``lateral deviation'' distance parameter. This is defined as the absolute value of lateral deviation from a straight line between the initial and final locations of the flier, normalized by the length of that distance. Essentially, it is a measure of the additional flight distance that is due to navigation decisions alone. For instance, a straightforward navigation model that could be defined as ``fly upwind after a first detection of odor'' would have a zero lateral deviation. At the same time, the success rate of such a navigator would be the chance of a randomly selected flier to start initially at a lateral position that is smaller than the radius of a successful flight. The lateral deviation of all the successful flights presented in Fig.~\ref{fig:deviations} summarizes the results in the form of probability density functions; the respective averages and standard deviations for different turbulent intensities and for the single radius of 0.25 m. The results for other radii are similar to the presented one. As depicted from Fig.~\ref{fig:deviations}, it appears that the ``cost of search'' of the presented algorithm is relatively constant for different turbulent intensity values. Following a certain increase in the average and decrease of the standard deviation from 0.05 to 0.1, the values seem to be constant. We conclude that the proposed navigation model is both realizable (there are about 80\% of successful flights for randomly distributed fliers) and robust algorithm since it does not increase additional lateral counter-turns for increased turbulence. 
Additional parameters that provide metric evaluation to the model such as the flight time, search time (an accumulated time between the upwind flight portions) and the length of the flight trajectory are presented in table~\ref{tab:Success-rates} for the sake of completeness.

%

\subsection*{Meandering and gusts}

\begin{figure}[!ht]
\centering\includegraphics[width=1\textwidth]{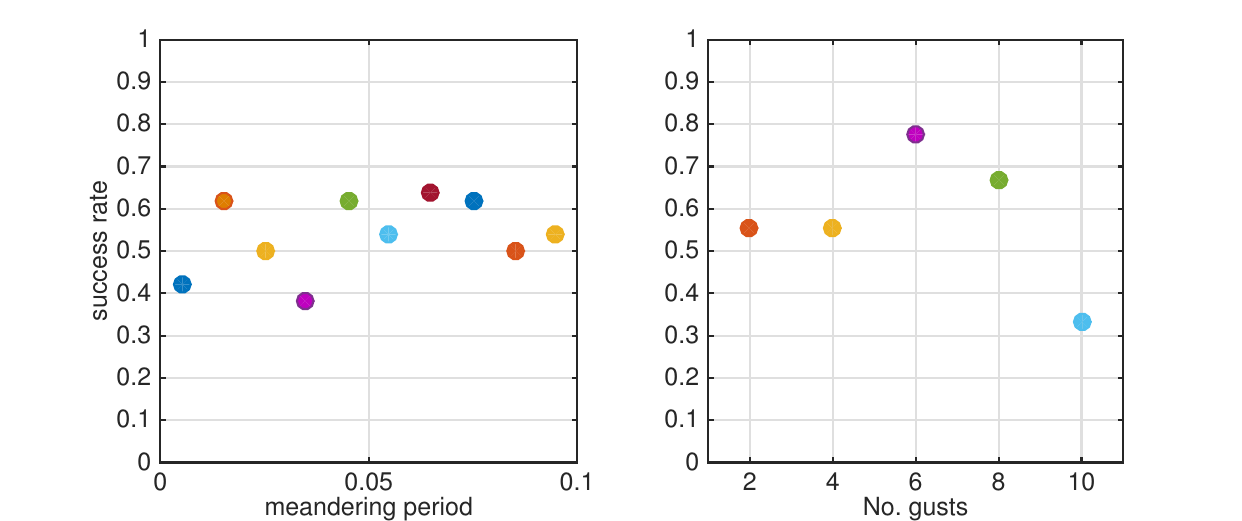} 
\caption{\added{{\bf Effect of meandering and gusts on search success} for different values of meandering or gusts shown in the legend. Graphs shown the average values.} \label{fig:meandering_gusts} }
\end{figure}

\added{Meandering is responsible for the slow drift of the mean wind direction. The main effect is in the decoupling of the direction of puffs motion at the release point from their general direction at later times. Optomotor anemotaxis mechanism assists the pheromone searching moth in its flight forward against the wind.  However, due to wind meandering,  wind direction does not provide valuable information regarding  the  location of the pheromone source. Thus males that rely on optomotor anemotaxis to reach a volatile source may not reach their target. A possible solution can rely on the adjustment of a moving direction with respect to the local wind conditions along the navigation path, or the re-acquisition of a plume using wide zigzagging in the cross-wind direction (casting). The width of the casting search shall be larger than the shift of the plume due to meandering. Using our navigation strategy is no different and success rate reduced for strong meandering, as shown in Fig.~\ref{fig:meandering_gusts}a.}  

\added{Gusts, on the contrary, have a minor effect on the success rate of the fliers that move according to the proposed strategy, as shown in Fig.~\ref{fig:meandering_gusts}b. This is because of the averaging effect along the puff trajectories and the notion that moth can turn following counter-turning at the distances that are limited by the product of $t_c$ and the constant ground speed. In some occasions the simulation is wrong; this occur in places where puffs that belong to the different groups of puffs interfere, before and right after the gust, but not in most random events of detection by a moving flier. }


\subsection*{Model sensitivity to internal counter concept}
\begin{figure}[!ht]
\centering\includegraphics[width=.8\textwidth]{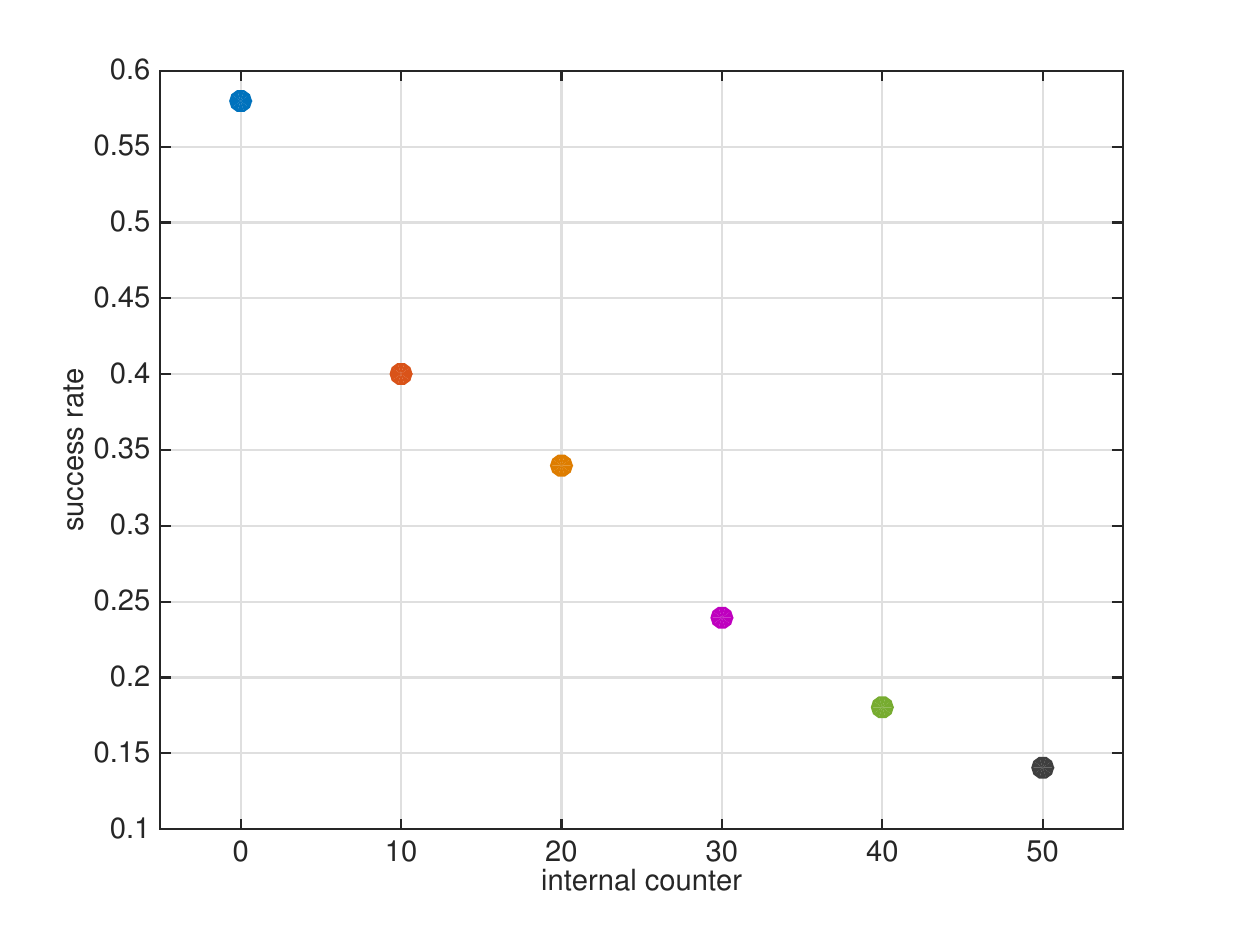} 
\caption{\added{{\bf Success rate for a navigation strategy with an internal counter} for different values of the counter. The point at zero is the success rate of a proposed navigation strategy with dynamically counter equal to $t_c$.} \label{fig:internal_counter} }
\end{figure}

\added{The last test is a sensitivity test to the proposed strategy in which counter-turning and upwind surges are proportional to the recent history of pheromone puffs detection that includes upwind propagation upon detection with a fixed ``internal counter'' controlled counter-turning. The results are shown in Fig.~\ref{fig:internal_counter} for a set of chosen internal counter values. The flier maintains a constant ground speed and a constant counter-turning time of flight. This creates a constant width of the zigzagging motion along a plume, alike some of the strategies proposed~\cite{Willis1991,Li2001,Bau2015} among others. The comparison of the results in Fig.~\ref{fig:internal_counter} with the results shown above reveals that in our case, internal counter approach may not be appropriate: without some additional modulation, the constant zigzagging width assumption will fail at any variable which is not tuned to the actual width of a ``virtual plume''. It means that this navigation strategy is not robust and can succeed only with the addition of wider than a plume casting or another re-acquisition option~\cite{Bau2015}. } 


\section*{Summary and conclusions}

In this study we have developed a bio-inspired algorithm for the navigation of a self-propelled flier towards a pulsating source of a scalar, convected in a turbulent flow. The algorithm could equally apply to any fluid, for instance for water convected concentration of a fluorescent dye and the navigation of a self-propelled swimmer, however in the present work we focused on parameters relevant for airborne plumes of odor. The flier sensory system is assumed to be simple and can only detect the presence of odor above a certain concentration level. 

\added{The model does not require a stereoscopic (two antennae) or additional (e.g. wind velocity and direction) measurement sensory system, neither previous knowledge or information obtained during the previous search history.}
We built the model based on the physics of a plume from a pulsating source, in which the size and distance between the puffs are related to each other through properties of a turbulent flow. Therefore, it is possible that the size of the puffs provides the flier with sufficient information in order to robustly locate the source in this complex environment with ample turbulence intensity. We modeled the puffs using a Lagrangian model in Eq.~\ref{eq:series}. The proposed model is contingent upon a number of biotic/abiotic variables that affect the probability of success of the flier to locate the source, among which the main quantity is the turbulent intensity. We then utilized the simulated navigation algorithm by releasing numerous randomly positioned virtual fliers in randomly generated turbulent fields of odor parcels, at different levels of turbulent intensity, meandering and gusts, and study their trajectories and resulting performance metrics. With respect to turbulence intensity we analyzed the success rates, search time, and the total flight time (as a measure of energy expenditure). We can conclude the effect of turbulent intensity as follows:
\begin{enumerate}
\item Both number of search cycles (each cycle can contain several counter-turns), and the number of counter-turns are reduced under increasing turbulence intensity. This is explained due to the larger area of the pheromone patches, and the consequent higher rate of flier interaction with detectable patches of odor. 
\item The number of times a flier crosses filaments of odor is greater when high turbulence conditions exist because of more frequent changes in the wind direction and broader distribution of patches. 
\item The total flight time increases when turbulence is increased; therefore, it can become larger than the time permitted for a navigator to find the source of odor and result in search termination.   
\item The probability of a successful search does not depend significantly on either $\alpha_{smin}$ or $\alpha_{smax}$ (the limits of the counter-turning angle). 
\end{enumerate}

%

The probability of the successful attempts of the flier to locate the source of the odor is about 80\% for the cases with varying turbulent intensity. This quantity does not seem to depend strongly on the turbulent intensity adopted in our simulations and is close to the probability of the successful events observed in relevant field experiments with some moths (see for example \cite{Carde2008}). Regarding meandering and gusts, we found that meandering slow drift of the mean velocity direction reduces the success rates of the fliers navigating with the proposed strategy. Nevertheless, a comparison with the fixed internal counter strategy demonstrates that without a wide casting component, our strategy is still more efficient and robust than the fixed counter-turning cycles approach. 

We conclude that the agreement of the model outlined here with the previously reported experimental findings lends credence to the hypothesis that a time duration of a detected event (crossing an odor patch) can serve as a viable cue for moth navigation. We also demonstrate that it is possible to construct a simple feasible and robust navigation model for a self-propelled flier, based on a single measurable quantity without relying on concentration of odor from the pulsed source. Additional study is necessary to apply the suggested moth-inspired algorithm of search to controlled fliers designed to detect and to locate the pulsating source of a chemical substance.

\added{Since most insect motion and communication in nature takes place in turbulent conditions \cite{Strand:2009gl,Thistle:2004vj}, and some species release pheromone as pulses, the present algorithm can shed light on an ecologically relevant model of navigation.}

\section*{Acknowledgments}
This study is partially supported by the U.S.-Israel Binational Science Foundation (BSF) under grant 2013399. 


\newpage{}
\section*{Tables}

\begin{table}[!ht]
\caption{{\bf Summary of simulation results (average $\langle x \rangle$ and standard deviation $S_x$) for the three levels of turbulent intensity: trajectory
length, time of flight , lateral
deviations, number of search cycles and number of counter-turns} The 0.05 (low) turbulence
intensity pertains to field or wind tunnel conditions, 0.15 to a typical value in open field atmospheric boundary layer, and the 0.3 (high)
turbulent intensity pertains to the air conditions in dense canopy
layers, for example, in forests. \label{tab:Success-rates} } 
\begin{center}
\begin{tabular}{|l|c|c|c|c|c|c|}
\hline 
Parameter  & \multicolumn{2}{c|}{0.05} & \multicolumn{2}{c|}{0.15}& \multicolumn{2}{c|}{0.3} \tabularnewline
\hline  
 & $\langle x \rangle $  & $S_x$  & $\langle x \rangle $  & $S_x$ & $\langle x \rangle $  & $S_x$ \tabularnewline
\hline 
Trajectory length (m)  & 9.68 & 1.82 & 10.14 & 2.17 & 9.88 & 2.14 \tabularnewline \hline 
Total flight time (s)  & 73.37 & 25.62 & 64.44 & 22.46 & 51.59 & 18.79 \tabularnewline \hline 
Search time (s)  & 14.13 & 6.05 & 18.03 & 5.94 & 22.86 & 4.95 \tabularnewline \hline 
Transverse deviation (-)  & 2.53 & 1.55 & 5.03 & 3.01 & 4.08 & 3.05 \tabularnewline \hline 
\# search cycles  & 26.39 & 11.25 & 25.09 & 8.73 & 16.78 & 4.97 \tabularnewline \hline 
\# counter-turns  & 
\multicolumn{2}{c|}{28.49} & \multicolumn{2}{c|}{17.74}& \multicolumn{2}{c|}{13.28} \tabularnewline \hline
\end{tabular}
\end{center}
\end{table}

\newpage{}


\section*{Figure legends}

\renewcommand{\thefigure}{B\arabic{figure}}
\setcounter{figure}{0}

\end{document}